\documentclass[reqno,11pt]{article}
\RequirePackage{amsthm, amsmath, amsfonts, natbib, graphicx}
\usepackage{supertabular}
\usepackage{bm}
\usepackage{amsmath}

\usepackage{booktabs}
\newtheorem{theorem}{Theorem}[section]

\usepackage{multirow}
\usepackage{array}
\usepackage{algorithm}
\usepackage{subfig}
\usepackage{algorithmic}
\usepackage{hyperref}
\hypersetup{colorlinks=true,citecolor=blue}

\theoremstyle{plain}

\theoremstyle{remark}

\newtheoremstyle{assm} 
{\topsep}                    
{\topsep}                    
{}                               
{}                           
{\scshape}                   
{.}                          
{.5em}                       
{}  
\theoremstyle{assm}

\theoremstyle{definition}

\newcommand{\ignore}[1]{}

\newcommand{\btheta}{\bm{\theta}}

\title{Uncertainty Quantification in Ensembles of Honest Regression Trees using Generalized Fiducial Inference}
\author{Suofei Wu\footnote{{\tt swu@ucdavis.edu}, Department of Statistics, University of California at Davis}
\and Jan Hannig \footnote{{\tt jan.hannig@unc.edu}, Department of Statistics \& Operations Research, University of North Carolina at Chapel Hill}
\and Thomas C. M. Lee\footnote{{\tt tcmlee@ucdavis.edu}, Department of Statistics, University of California at Davis}
}
\date{November 10, 2019}

\begin{document}        
\maketitle
\begin{abstract}
Due to their accuracies, methods based on ensembles of regression trees are a popular approach for making predictions.  Some common examples include Bayesian additive regression trees, boosting and random forests.  This paper focuses on honest random forests, which add honesty to the original form of random forests and are proved to have better statistical properties.  The main contribution is a new method that quantifies the uncertainties of the estimates and predictions produced by honest random forests.  The proposed method is based on the generalized fiducial methodology, and provides a fiducial density function that measures how likely each single honest tree is the true model.  With such a density function, estimates and predictions, as well as their confidence/prediction intervals, can be obtained.  The promising empirical properties of the proposed method are demonstrated by numerical comparisons with several state-of-the-art methods, and by applications to a few real data sets.  Lastly, the proposed method is theoretically backed up by a strong asymptotic guarantee.
\end{abstract}
{\bf Keywords:} additive regression trees, confidence intervals, FART, prediction intervals, random forests

\section{Introduction}
Ensemble learning is a popular method in regression and classification because of its robustness and accuracy \citep{mendes2012ensemble}.  It is commonly used to make predictions for future observations.  Denote the observed sample as $\{Y_i,\bm{X}_i\}$, $i=1,\ldots,n$, where $Y_i\in \mathbb{R}$ are scalar responses and $X_i \in \mathbb{R}^p$ are vector predictors. The general regression model is 
\begin{equation}
\label{regression_problem}
Y_i=f(\bm{X}_i)+\epsilon_i,
\end{equation}
where the iid noise $\epsilon_i$'s follow $N(0,\sigma^2)$.  An ensemble learning method approximates the model $f(\cdot)$ by a weighted sum of weak learners $T_i(\cdot)$'s with weights $w_i$'s:
\begin{equation}\label{ensemble}
f(\bm{X}_i)=\sum_{i=1}^a w_iT_i(\bm{X}_i).
\end{equation}

Decision tree is a common choice for the weak learners because it has high accuracy and flexibility.  Although it suffers from high variance, an ensemble of decision trees will keep the accuracy and at the same time reduce the variance.  Given their successes, ensembling of trees have attracted a lot of attention.  For example, Random forests \citep{breiman2001random} and bagging \citep{breiman1996bagging} take average of decorrelated trees to obtain a more stable model with similar bias.  While both bagging and random forests sample a different training set with replacement when growing each tree, random forests also consider a randomly selected subset of features for each split.  Recently, \citet{athey2019generalized} proposed generalized random forests that construct a more general framework and can be naturally extended to other statistical tasks such as quantile regression and heterogeneous treatment effect estimation.  All of the three methods use the basic ensemble method (BEM) \citep{perrone1992networks} in regression, which takes all the $w_i$'s in \eqref{ensemble} equally as ${1}/{a}$.

\citet{wang2003mining} proposed a weighted ensemble approach for classification, where the classifiers are weighted by their accuracies in classifying their own training data.  Their work can be straightforwardly extended to the regression case.  Bayesian ensemble learning \citep{wu2007bayesian,chipman2007bayesian} is another approach that takes a weighted average of the trees.  The posterior probabilities are used as weights in this scenario.

\begin{figure}[ht]
\begin{center}
\includegraphics[scale=0.3]{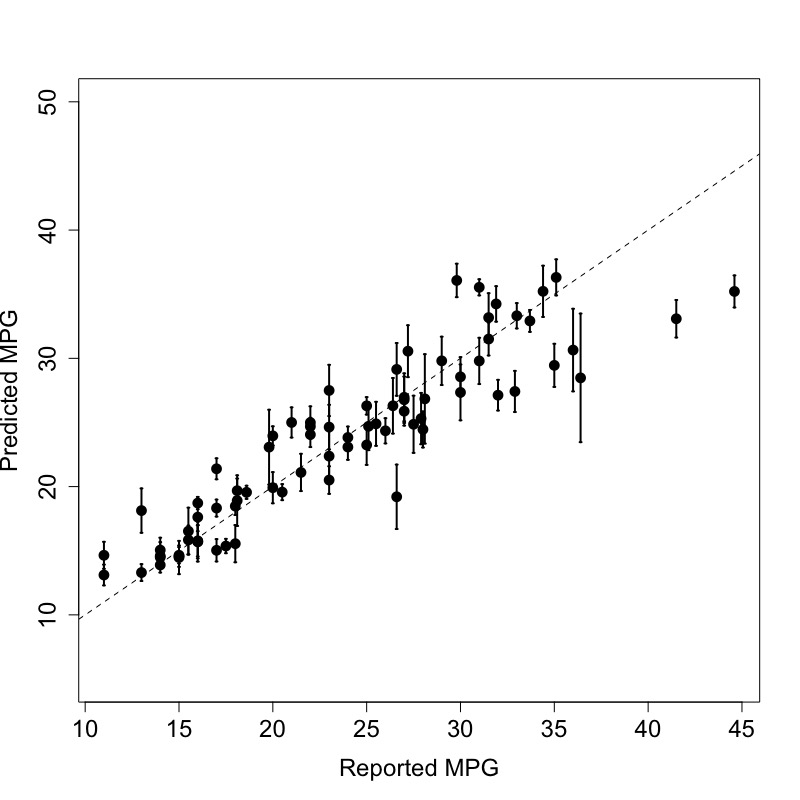}	
\caption{Random forest predictions and $95\%$ confidence intervals of the \emph{Auto MPG} data set.}
\label{auto_mpg}
\end{center}
\end{figure}

Despite the above efforts, the study of uncertainty quantification of ensemble learning is somewhat limited.  One notable exception is \citet{wager2014confidence}, where the authors proposed a method that produces standard error estimates $\hat{\sigma}$ for random forests predictions.  It is based on jackknife and infinite jackknife \citep{efron2014estimation} and can be used for constructing Gaussian confidence intervals.  Figure~\ref{auto_mpg} shows the result of applying their method on the \emph{Auto MPG} data set.  The goal is to predict fuel economy of automotives (in miles per gallon, MPG) using $7$ features.  Further details of this data set can be found in Section~\ref{sec_real_data} below. Following \citet{wager2014confidence}, we randomly split the data into a training set of size $314$ and a testing set of size $78$. The error bars in Figure~\ref{auto_mpg} are $1.96$ standard error in each directions. The rate that these error bars cover the prediction-equals-observation diagonal is $46\%$.  This suggests that there are residual noise in the data that cannot be explained by the random forests model based on the available features.

Moreover, in simulation experiments, the confidence intervals do not have a good coverage rate either, especially when there are a lot of noise in $\bm{X}$.  We repeat the same simulation setting as in \citep{wager2014confidence} and report the results in Section~\ref{c3_simulation}.

Besides the work of \citet{wager2014confidence}, \citet{mentch2016quantifying} showed that under some strong assumptions, random forests based on subsampling are asymptotically normal, allowing for confidence intervals to accompany predictions.  In addition, \citet{chipman2010bart} developed a Bayesian Additive Regression Trees model (BART) that produces both point and interval estimates via posterior inference.

In this paper, we use generalized fiducial inference \citep{hannig2016generalized} to construct a probability density function on the set of honest trees in an honest random forests model.  We shall show that such a new ensemble method of honest trees provides more precise confidence intervals as well as point estimates.

The rest of this paper is organized as follows.  First a brief introduction of generalized fiducial inference is provided in Section~\ref{sec:GFIbackground}.  Then the main methodology is presented in Section~\ref{sec:method} and the theoretical properties of the method is studied in Section~\ref{sec:theory}.  Section~\ref{c3_simulation} illusrates the practical performances of the proposed method.  Lastly concluding remarks are offered in Section~\ref{sec:conclude} while technical details are delayed to the appendix.

\section{Generalized Fiducial Inference}
\label{sec:GFIbackground}
Fiducial inference was first introduced by Fisher in \citep{fisher1930inverse}.  It aims to construct a statistical distribution for the parameter space when no prior information is available.  Under such condition, the usage of classical Bayesian framework receives criticism because it requires a prior distribution of the parameter space.  Alternatively, Fisher considered a switching mechanism between the parameters and the observations, which is quite similar to how parameters are estimated by the maximum likelihood method.  Despite Fisher's continuous effort on the theory of fiducial inference, this framework was overlooked by the majority of the statistics community for several decades.  \citet{hannig2016generalized} has a detailed introduction on the history of the original fiducial inference.

In recent years, there is a renewed interest in extending Fisher's idea.  The modified versions include Dempster-Shafer theory \citep{dempster2008dempster, martin2010dempster}, inferential models \citep{martin2015conditional, martin2015marginal}, confidence distributions \citep{xie2013confidence,xie2011confidence} and generalized inference \citep{weerahandi1995generalized, weerahandi2013exact}.  In this paper we focus on the successful extension known as generalized fiducial inference (GFI) \citep{hannig2006fiducial}.  It has been successfuly applied to a variety of problems, including wavelet regression \citep{hannig2009wavelet}, ultrahigh-dimensional regression \citep{randy2015}, logistic regression \citep{liu2016generalized} and nonparametric additive models \citep{Gao-etal19}.

Under the GFI framework, the relationship between the data $\bm{y}$ and the parameter $\bm{\theta}$ is expressed by a data generating equation $G$:
\begin{equation*}
\bm{y}=G(\bm{u},\btheta),
\end{equation*}
where $\bm{u}$ is a random component with a known distribution.  Suppose for the moment that the inverse function $G^{-1}$ exists for any $\bm{u}$; i.e., one can always calculate $\bm{\theta}=G^{-1}(\bm{u},\bm{y})$ for any $\bm{u}$.  Then a random sample $\{\tilde{\bm{\theta}}_1, \tilde{\bm{\theta}}_2, \ldots\}$ of $\bm{\theta}$ can be obtained by first generating a random sample $\{\tilde{\bm{u}}_1, \tilde{\bm{u}}_2, \ldots\}$ of $\bm{u}$ and then calculate
\begin{equation*}
\tilde{\bm{\theta}}_1=G^{-1}(\tilde{\bm{u}}_1, \bm{y}), \quad \tilde{\bm{\theta}}_2=G^{-1}(\tilde{\bm{u}}_2, \bm{y}), \quad \ldots
\end{equation*}
Notice that the roles of $\bm{\theta}$ and $\bm{u}$ are ``switched'' in the above, as in the maximum likelihood method of Fisher.  See \citet{hannig2016generalized} for strategies to ensure the existence of $G^{-1}$.  We call the above random sample $\{\tilde{\bm{\theta}}_1, \tilde{\bm{\theta}}_2, \ldots\}$ a {\em generalized fiducial sample} of $\bm{\theta}$ and the corresponding density $r(\bm{\theta})$ the {\em generalized fiducial density} of $\bm{\theta}$. 

Beyond this conceptually appealing and well defined definition, \citet{hannig2016generalized} provides a user friendly formula for the fiducial density
\begin{equation}
r(\btheta)=\frac{h(\bm{{y}},\btheta)J(\bm{{y}},\btheta)}{\int_\Theta h(\bm{{y}},\btheta^{'})J(\bm{{y}},\btheta^{'})d\btheta^{'}},
\label{fiducial_density}
\end{equation}
where $h(\bm{{y}},\btheta)$ is the likelihood and the function
\begin{equation*}
J(\bm{{y}},\btheta)=D\left\{\nabla_{\bm{\theta}}\bm{G(\bm{u},\btheta)}|_{\bm{{u}}=\bm{G}^{-1}(\bm{{y},\btheta})}\right\}
\end{equation*}
with $D(\bm{A})=\{\det(\bm{A}^T\bm{A})\}^\frac{1}{2}$.

Formula~(\ref{fiducial_density}) assumes that the model dimension is known.  When model selection is involved, the generating function of a certain model $T$ becomes:
\begin{equation*}
\bm{y}=G(\btheta_T,\bm{u},T).
\end{equation*}
Similar to maximum likelihood, GFI tends to assign higher probabilities to models with higher complexity (i.e., larger number of parameters).  As similar to penalized maximum likelihood, \citet{hannig2009wavelet} suggested adding an extra penalty term to~(\ref{fiducial_density}).  The marginal fiducial probability of a specific model $T$ then becomes:
\begin{equation}\label{marginal_density}
r(T)=\frac{\int r_T(\btheta_T)n^{-\frac{l(T)}{2}}d\btheta_T}{\sum_{T^{'}\in\mathcal{T}}\int r_{T^{'}}(\btheta_{T^{'}})n^{-\frac{l(T^{'})}{2}}d\btheta_{T^{'}}},
\end{equation} 
where $\mathcal{T}$ is the set of all possible models and $l(T)$ is the number of parameters in model $T$; see \cite{hannig2016generalized} for derivation.  Therefore in practice, when model selection is involved, to generate a fiducial sample for $\btheta_T$, one can first choose a model $\tilde{T}$ using~(\ref{marginal_density}), and then select $\tilde{\btheta}_T$ from~(\ref{fiducial_density}) given $\tilde{T}$.  We note that closed form expressions for~(\ref{fiducial_density}) and~(\ref{marginal_density}) do not always exist so one may need to resort to MCMC techniques.

\section{Methodology}
\label{sec:method}
\subsection{Regression Trees and Honest Regression Trees}
A decision tree models the function $f(\cdot)$ in~\eqref{regression_problem} by recursively partitioning the feature space (i.e., the space of all $\bm{X}$'s) into different subsets.  These subsets are called {\em leaves}.  Let $\bm{X}_0$ be any point in the feature space and $L(\bm{X}_0)$ be the leaf that contains $\bm{X}_0$.  The decision tree estimate $\hat{f}(\bm{X}_0)$ for $f(\bm{X}_0)$ is the average of those responses that are in the same leaf as $\bm{X}_0$:
\begin{equation*}
\hat{f}(\bm{X})=\frac{1}{|\{i:\bm{X}_i\in L(\bm{X})\}|}\sum_{i:\bm{X}_i\in L(\bm{X})}Y_i.
\end{equation*}
Naturally one may like a partition that minimizes the loss function:
\begin{equation*}
\sum_{i=1}^n\{y_i-\hat{f}(\bm{X}_i)\}^2.
\end{equation*}
However, very often in practice a serious drawback is that the number of potential partitions is huge which makes it infeasible to obtain the partition that minimizes the above loss.  Therefore, a greedy search algorithm is usually considered, which consists of the following steps:
\begin{enumerate}
\item Start from the root.
\item Choose the best feature and split point that minimize the loss function.
\item Recursively repeat the former step on the children nodes.
\item Stop when 
\begin{itemize}
\item Each node achieves the minimum node size pre-specified by the user, or
\item The loss cannot be further reduced by extra partitioning.
\end{itemize}
\end{enumerate}

One criticism about the above decision tree is that the same data are used to grow the tree and make prediction.  To ensure good statistical behaviors and as a response to this criticism, honest decision trees were proposed \citep{biau2012analysis, denil2014narrowing}.  An honest tree is grown using one subsample of the training data while uses a different subsample for making predictions at its leaves.  If there are no observations falling to a specific leaf, its prediction will be made by one of its parents.  A corresponding honest random forest can be generated by using the same mechanism to generate random forests from decision trees.  \citet{wager2018estimation} proved that under some regularity conditions, the leaves of an honest tree become small in all dimensions of the feature space when $n$ becomes large.  Hence, if we also assume that the true generating function is Lipschitz continuous, honest trees are unbiased and so are honest random forests.

\subsection{Ensemble of Honest Trees using Generalized Fiducial Inference}
The goal is to solve the regression problem (\ref{regression_problem}) using an ensemble of honest trees $\{T_j\}_{j=1}^a$ and apply GFI to conduct statistical inference.

\begin{figure}
	\begin{center}
		\includegraphics[scale=0.8]{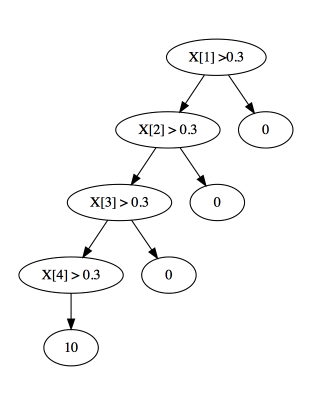}	
		\caption{A function $T$ that has a binary tree structure}
		\label{tree}
	\end{center}
\end{figure}

Suppose there exists a binary tree structured function $T_0$ such that $f(\bm{X})=T_0(\bm{X})$ for any $\bm{X}\in \mathbb{R}^P$; we will call any such tree a {\em true model}. One example is the ``AND'' function mentioned in \citet{wager2014confidence}:
\begin{equation*}
Y=10\times \text{AND}(X_1>0.3; X_2>0.3; X_3>0.3; X_4>0.3)+\epsilon.
\end{equation*}
A corresponding binary tree $T$ is shown in Figure~\ref{tree}.
	
We want to assign a generalized fiducial probability to each tree $T$ measuring how likely the true generating function is contained in $T$.  Suppose $T$ has ${l(T)}$ leaves, $L_1,\ldots, L_{l(T)}$.  Denote the number of observations in the $j$-th leaf is $n_j$ and hence $n=n_1+\cdots+n_{l(T)}$.  Also denote the response value of the $j$-th leaf as $\mu_j$, which can be estimated by the average of all the $Y_i$'s that belong to this leaf:
\begin{equation*}
\hat{\mu}_{j}=\frac{1}{|L_j|}\sum_{i:\bm{X}_i\in L_j}Y_i.
\end{equation*}
Write $\bm{\mu}=(\mu_1,...,\mu_k)$.  First we calculate the generalized fiducial density $r(\bm{\mu}, \sigma^2)$ for $\bm{\theta}_T=\{\bm{\mu}, \sigma^2\}$.  From~(\ref{fiducial_density}) it can be shown that $r(\bm{\mu}, \sigma^2)$ is proportional to
\begin{equation}
\label{density_function}
r(\bm{\mu},\sigma^2) \propto
J(\bm{\mu},\sigma^2)\frac{1}{(2\pi\sigma^2)^\frac{n}{2}}e^{\frac{-\sum(Y_i-\mu_{i'})^2}{2\sigma^2}}n^{-\frac{{l(T)}}{2}},
\end{equation}
where $i'$ is the index of the leaf that $\bm{X}_i$ belongs to; i.e., $\bm{X}_i\in L_{i'}$.

The Jacobian term in (\ref{density_function}) is
\begin{equation*}
J(\bm{\mu},\sigma^2)=\frac{\sqrt{\text{SSE}\prod n_i}}{\sigma}
\end{equation*}
with $\text{SSE}=\sum(Y_i-\hat{Y}_i)^2$ as the sum of squared errors, where $\hat{Y}_i =\hat{\mu}_{i'}$, the average of all $Y_i$'s belong to the leaf $L_{i'}$.

Now we can calculate the marginal fiducial density for the tree $T$ using~(\ref{marginal_density}), for which the numerator becomes:
\begin{eqnarray}
r(T)
& \propto & \int\int\frac{\sqrt{\prod n_i \text{SSE}}}{\sigma}(\frac{1}{\sqrt{2\pi}\sigma})^ne^{\frac{-\sum(Y_i-\bar{Y}_i+\bar{Y}_i-\mu_i)^2}{2\sigma^2}}n^{-\frac{{l(T)}}{2}} d\mu_1,\ldots,d\mu_kd\sigma^2 \nonumber \\
&=& \int\int\frac{\sqrt{\prod n_i \text{SSE}}}{\sigma}(\frac{1}{\sqrt{2\pi}\sigma})^ne^{\frac{-\sum(Y_i-\bar{Y}_i)^2}{2\sigma^2}-\frac{\sum(\bar{Y}_i-\mu_i)^2}{2\sigma^2}}n^{-\frac{{l(T)}}{2}} d\mu_1,\ldots,d\mu_kd\sigma^2 \nonumber \\
&=&\int\frac{\sqrt{\text{SSE}}}{(2\pi)^{\frac{n-{l(T)}}{2}}\sigma^{n-{l(T)}+1}}e^{-\frac{\text{SSE}}{2\sigma^2}}n^{-\frac{{l(T)}}{2}}d\sigma^2 \nonumber \\
&=&\int \text{SSE}^{\frac{1}{2}-\frac{n-{l(T)}+1}{2}+1}2^{-\frac{1}{2}}\pi^{-\frac{n-{l(T)}}{2}}\xi^{\frac{n-{l(T)}+1}{2}-2}e^{-\xi}n^{-\frac{{l(T)}}{2}}d\xi \nonumber \\
&\propto & \frac{\Gamma(\frac{n-{l(T)}-1}{2})n^{-\frac{{l(T)}}{2}}}{\text{SSE}^{\frac{n-{l(T)}}{2}-1}\pi^{\frac{n-{l(T)}}{2}}}.
\label{eqn:RT}
\end{eqnarray}

\subsection{A Practical Method for Generating Fiducial Samples}
This subsection presents a practical method for generating a fiducial sample of honest trees using~(\ref{eqn:RT}).

Even when $n$ is only of moderate size, the set $\mathcal{T}$ of all possible trees is huge and therefore we only consider a subset of trees $\mathcal{T}^* \subset \mathcal{T}$.  More precisely, $\mathcal{T}^*$ is an honest random forest with an adequate number of trees such that one can assume it contains at least one true model $T$.  Each tree samples $\lfloor n/4\rfloor$ observations without replacement to grow, and uses a different group of $\lfloor n/4\rfloor$ observations to calculate the averages $\hat{\mu}_j$'s (i.e., make predictions) at the leaves.

Loosely, three steps are involved in generating a fiducial sample of trees.  The first step is to generate the structure of the tree, the second step is to generate the noise variance, and the last step is to generate the leaf values $\mu_j$'s.  We begin with approximating the generalized fiducial density $r(T)$ in~(\ref{eqn:RT}) as follows.  

For each tree $T\in \mathcal{T}^*$, we calculate:
\begin{equation*}
R(T)=\frac{\Gamma(\frac{n-{l(T)}-1}{2})n^{\frac{-{l(T)}}{2}}}{\text{SSE}^{\frac{n-{l(T)}}{2}-1}\pi^{\frac{n-{l(T)}}{2}}}
\end{equation*}
and approximate $r(T)$ with
\begin{equation}\label{density}
r(T)=\frac{R(T)}{\sum_{T^{'}\in\mathcal{T}^*}R(T^{'})}.
\end{equation}
After a tree $T$ is sampled from (\ref{density}), $\tilde{\sigma}^2$ is sampled from
\begin{equation}\label{sigma}
\text{SSE}/\sigma^2\sim\chi^2_{n-{l(T)}},
\end{equation}
where ${l(T)}$ denotes the number of leaves in $T$.  To sample the $\mu_j$'s, we draw without replacement $\lfloor n/4\rfloor$ observations from the part of the data that was not used to grow $T$.  Denote these drawn observations as $\{\bm{X}_i^*, Y_i^*\}_{i=1}^{\lfloor n/4\rfloor}$.  Then a generalized fiducial tree sample $\tilde{T}$ can be obtained by updating the leaf values of $T$ using
\begin{equation}
\tilde{\mu}_{j}=\frac{1}{|L_j|}\sum_{i:\bm{X}_i^*\in L_j}Y_i^*+\tilde{\sigma} z_i,
\label{eqn:newmu}
\end{equation}
where $z_l \overset{\text{iid}}{\sim} N(0,1)$, $j=1, \ldots, {l(T)}$ with ${l(T)}$ being the number of leaves in $T$.

Repeating the above procedure multiple times provides multiple copies of the fiducial sample $\{\tilde{\sigma},\tilde{T}\}$.  Statistical inference can then be conducted in a similar fashion as with a posterior sample in the Bayesian context.  For any design point $\bm{X}$, averaging over all the $\tilde{T}(X)$'s will deliver a point estimate for $f(\bm{X})$.  The $\alpha/2$ and $1-\alpha/2$ percentiles of $\tilde{T}(\bm{X})$ will give a $100(1-\alpha)\%$ confidence interval for $f(\bm{X})$, while the $\alpha/2$ and $1-\alpha/2$ percentiles of $\tilde{T}(\bm{X})+\tilde{\sigma}z$ will provide a prediction interval for the corresponding $Y$.

We summarize the above procedure in Algorithm~\ref{a1}.
\begin{algorithm}
\caption{A Generalized Fiducial Method for Generating Honest Tree Ensemble}\label{a1}
\begin{algorithmic}[1]
\STATE Choose $N_{T}$ and $M$.
\STATE Train an honest random forest with $N_{T}$ honest trees $\{T_j\}_{j=1}^{N_{T}}$.
\STATE For each tree $T_j$, calculate the generalized fiducial probability $r(T_j)$ using~(\ref{density}).
\FOR {$i=1, \ldots, M$}
\STATE Draw a $T\in \{T_j\}_{j=1}^{N_{T}}$ using (\ref{density}).
\STATE Draw a $\tilde{\sigma}^2$ from (\ref{sigma}).
\STATE Draw without replacement $\lfloor n/4\rfloor$ observations from the part of the data that was not used to grow $T$.  Denote these drawn observations as $\{\bm{X}_i^*, Y_i^*\}_{i=1}^{\lfloor n/4\rfloor}$.
\STATE Obtain $\tilde{T}$ by updating its leaf values using (\ref{eqn:newmu}).
\ENDFOR
\STATE Output the $M$ copies of generalized fiducal sample $\{\tilde{\sigma}^2, \tilde{T}\}$ obtained from above for further inference.
\end{algorithmic}
\end{algorithm}

\section{Asymptotic Properties}
\label{sec:theory}


The theoretical properties of the proposed method is established under the following conditions:

A1) The generating function $f(x)$ has a binary tree structure. Denote the training data set of $T$ as $\mathcal{D}_T$, ~$\mathcal{D}_T=\{Y_i,\bm{X}_i\}_{i=1}^{\lfloor n/4\rfloor}$. We say this binary tree $T$ is a true model, if for any $\bm{X}$ in the training set $\mathcal{D}_T$, $\mathbb{E}(T(\bm{X}))=f(\bm{X})$. Notice that such a binary tree is not unique. We denote the collections of true models as $\mathcal{T}_0$:
\begin{equation*}
\mathcal{T}_0=\{T:T \text{ is a true model}\}.
\end{equation*}

A2) Let $\mathcal{T}$ be the collection of honest trees in a trained random forests model. We assume that $\mathcal{T}$ should have at least one tree that belongs to $\mathcal{T}_0$:
\begin{equation*}
P(\mathcal{T}\cap\mathcal{T}_0=\emptyset)\rightarrow 0.
\end{equation*}

A3) Meanwhile, we assume that the size of $\mathcal{T}$ is not too large for practical use.
\begin{equation*}
|\mathcal{T}|=o(\sqrt{\frac{\log(n)}{\log\log n}}).
\end{equation*}

A4) Let $\bm{H}_T$ be the projection matrix of $T$; i.e., $\bm{H}_T=\{h_{ij}\}_{i,j=1}^n$, 
$$
h_{ij}=
    \begin{cases}
    \frac{1}{n_i}, \,\,\text{if}\,\,\bm{X_i}\in L(\bm{X_j}) \,\text{in}\, T\\
    0,  \,\,\text{else.}
    \end{cases}
$$

Let $\Delta_T=||\bm{\mu}-\bm{H}_T\bm{\mu}||^2$, where $\bm{\mu} =E(\bm{y})$.  Assume
	\begin{equation}
	\lim_{ n\rightarrow \infty}\min_{T\in\mathcal{T}\backslash\mathcal{T}_0}\left\{\frac{\Delta_T}{l(T)\log n} \right\} = \infty.
	\label{eqn:a1}
	\end{equation}

A5) Denote the number of leaves of a tree $T$ as $l(T)$. Let $l_0$ be the minimum number of leaves of the trees in $\mathcal{T}\cap\mathcal{T}_0$:
	\begin{equation*}
	l_0=min\{l(T),T\in\mathcal{T}\cap\mathcal{T}_0\}
	\end{equation*}
	and $\mathcal{T}_{l_0}$ be the trees in $\mathcal{T}\cap\mathcal{T}_0$ with number of leaves equals to $l_0$:
	\begin{equation*}
	\mathcal{T}_{l_0}=\{T:l(T)=l_0, T\in\mathcal{T}\cap\mathcal{T}_0\}.
	\end{equation*}
	
A6) Denote $L$ as the maximum number of leaves in $\mathcal{T}$: $$L=\max\{l(T): T\in \mathcal{T}\}.$$ Assume that $L$ is at most $n^{\alpha}$, with $0<\alpha<1$.\\
Under the above assumptions, we have
\begin{theorem}\label{t1}
	\begin{equation*}
	\sum_{T\in \mathcal{T}_{l_0}}r(T)\rightarrow_p 1.
	\end{equation*}
\end{theorem}
The proof can be found in the appendix.

\section{Empirical Properties}
\label{c3_simulation}
This section illustrates the practical performance of the above proposed method via a sequence of simulation experiments and real data applications.  We shall call the proposed method FART, short for Fiducial Additive Regression Trees.

\subsection{Simulation Experiments}
In our simulation experiments three test functions were used:
\begin{itemize}
  \label{and}
	\item Cosine: $3\cdot\cos(\pi\cdot(X_1 + X_2))$,
	\item XOR: $5\cdot\mbox{XOR} (X_1 > 0.6; X_2 > 0.6) + \mbox{XOR} (X_3 > 0.6; X_4 > 0.6)$,
	\item AND: $10\cdot\mbox{AND} (X_1 > 0.3; X_2 > 0.3; X_3 > 0.3; X_4 > 0.3)$.
\end{itemize}
The design points $\bm{X}_i$'s are iid $U(0,1)$ and the errors $\epsilon_i$'s are iid $N(0,1)$.  We tested different combinations of $n$ and $p$ (see below).  These experimental configurations have been used by previous authors \citep[e.g.,][]{chipman2010bart, wager2014confidence}.  The number of repetitions for each experimental configuration is 1000.

We applied FART to the simulated data and calculated the mean coverages of various confidence intervals.  We also applied the following three methods to obtain other confidence intervals:
\begin{itemize}
\item BART: Bayesian Additive Regression Trees of \citet{chipman2010bart},
\item Bootstrap: the bootstrap method of \citet{mentch2016quantifying}, and
\item Jackknife: the infinite jackknife method of \citet{wager2014confidence}.
\end{itemize}
Tables~\ref{y_1},~\ref{y_2} and~\ref{y_3} report the empirical coverage rates of the, respectively, 90\%, 95\% and 99\% confidence intervals produced by these methods for $E(Y^*|\bm{X}^*)$, where $(\bm{X}^*, Y^*)$ is a random future data point.

Overall FART provided quite good and stable coverages.  The performances of Bootstrap and Jackknife are somewhat disappointing.  The possible reasons are that in Jackknife the uncertainty of the residual noise was not taken into account, and that Bootstrap is, in general, not asymptotically unbiased, as argued in \citet{wager2018estimation}.  BART sometimes gave better results than FART.  However, for those cases where BART were better, results from FART were not far behind, but for some other cases, BART's results could be substantially worse than FART's.  Therefore it seems that FART is the prefered and safe method if one is targeting $E(Y^*|\bm{X}^*)$.


\begin{table}[htbp]
\centering
\caption{Empirical coverage rates for the $90\%$ confidence intervals for $E(Y^*|\bm{X}^*)$ obtained by the various methods.  The numbers in parentheses are the averaged widths of the intervals.  The results that are closest to the target coverage rate are highlighted in bold.}
\label{y_1}
\begin{tabular}{l|ll|llll}
\toprule
function & $n$ & $p$& FART & Bootstrap & Jackknife &BART \\
\midrule
Cosine & 50 & 2 & $\bm{82.7}$ $\bm{(4.29)}$ & 34.6 (1.63) & 23.6 (1.40) & 57.6 (2.33)  \\
Cosine & 200 & 2 & 87.4 (3.11) & 51.0 (2.12) & 39.2 (1.05) & $\bm{91.1}$ $\bm{(1.66) }$ \\
XOR & 50 & 50 & $\bm{73.2}$ $\bm{(4.64)}$ & 5.0 (1.72) & 3.8 (1.48) & 62.9 (4.53)  \\
XOR & 200 & 50 & 92.6 (2.61) & 26.3 (2.96) & 32.0 (1.02) & $\bm{89.1}$ $\bm{(3.53)}$  \\
AND & 50 & 500 & 60.3 (8.19) & 3.4 (3.07) & 0.7 (2.30) & $\bm{66.1}$ $\bm{(6.87)}$  \\
AND & 200 & 500 & $\bm{87.2}$ $\bm{(4.79)}$ & 35.0 (5.09) & 0.0 (1.94) & 59.2  (6.10)  \\
\bottomrule
\end{tabular}%
\end{table}

\begin{table}[htbp]
\centering
\caption{Similar to Table~\protect\ref{y_1} but for the $95\%$ confidence intervals of $E(Y^*|\bm{X}^*)$.}
\label{y_2}
\begin{tabular}{l|ll|llll}
\toprule
function & $n$ & $p$ & FART & Bootstrap & Jackknife &BART \\
\midrule
Cosine & 50 & 2 & $\bm{91.6}$ $\bm{(5.27)}$  & 41.9 (1.94) & 27.9 (1.67) & 66.3 (2.78)  \\
Cosine & 200 & 2 & 93.6 (3.80) & 58.8 (2.52) & 47.5 (1.26) & $\bm{95.6}$ $\bm{(1.98)}$   \\
XOR & 50 & 50 & $\bm{83.9}$ $\bm{(5.67)}$  & 8.0 (2.05) & 6.7 (1.77) & 77.3 (5.39)  \\
XOR & 200 & 50 & 96.1 (3.22) & 38.0 (3.53) & 37.4 (1.21) & $\bm{95.4}$ $\bm{(4.21)}$   \\
AND & 50 & 500 & $\bm{71.5}$ $\bm{(9.91)}$  & 8.3 (3.65) & 2.6 (2.74) & 71.4 (8.18)  \\
AND & 200 & 500 & $\bm{90.5}$ $\bm{(5.86)}$  & 50.3 (6.06) & 0.4 (2.31) & 67.9 (7.26)  \\
\bottomrule
\end{tabular}%
\end{table}

\begin{table}[htbp]
\centering
\caption{Similar to Table~\protect\ref{y_1} but for the $99\%$ confidence intervals of $E(Y^*|\bm{X}^*)$.}
\label{y_3}
\begin{tabular}{l|ll|llll}
\toprule
function & $n$ & $p$& FART & Bootstrap & Jackknife &BART \\
\midrule
Cosine & 50 & 2 & $\bm{98.0}$ $\bm{(7.30)}$ & 54.6 (2.55) & 39.0 (2.19) & 78.9  (3.64)  \\
Cosine & 200 & 2 & 98.6 (5.26) & 73.1 (3.32) & 60.5 (1.65) & $\bm{98.6}$ $\bm{(2.59)}$  \\
XOR & 50 & 50 & $\bm{95.1}$ $\bm{(7.81)}$ & 17.2 (2.69) & 13.4 (2.32) & 91.1  (7.03)  \\
XOR & 200 & 50 & $\bm{98.5}$ $\bm{(4.58)}$ & 62.9 (4.64) & 46.5 (1.60) & 98.2  (5.48)  \\
AND & 50 & 500 & $\bm{89.3}$ $\bm{(13.33)}$ & 24.9 (4.80) & 12.2 (3.60) & 75.3 (10.65)  \\
AND & 200 & 500 & $\bm{95.0}$ $\bm{(8.15)}$ & 69.0 (7.96) & 5.7 (3.04) & 81.7 (9.44)  \\
\bottomrule
\end{tabular}%
\end{table}

Next we examine the coverage rates for the noise standard deviation $\sigma$.  Since Bootstrap and Jackknife do not produce convenient confidence intervals for $\sigma$, we only focus on FART and BART.  The results are summarized in Tables~\ref{sigma_90}, \ref{sigma_95} and~\ref{sigma_99}.  Overall one can see that FART is the prefered method, although its performances for the test function AND were disappointing.

\begin{table}[htbp]
\centering
\caption{Empirical coverage rates for the $90\%$ confidence intervals for $\sigma$ obtained by FART and BART.  The numbers in parentheses are the averaged widths of the intervals.  The results that are closest to the target rate are highlighted in bold.}
\label{sigma_90}
\begin{tabular}{l|ll|ll}
\toprule
function & $n$ & $p$& FART  &BART \\
\midrule
Cosine & 50 & 2 & $\bm{93.4}$ $\bm{(0.68)}$&8.5 (0.67)\\
Cosine & 200 & 2 & $\bm{96.8}$ $\bm{(0.26)}$&86.5 (0.20)\\
XOR & 50 & 50 & $\bm{71.8}$ $\bm{(0.67)}$&5.4 (1.03)\\
XOR & 200 & 50 & $\bm{90.0}$ $\bm{(0.24)}$&93.6 (0.62)\\
AND & 50 & 500 & $\bm{24.3}$ $\bm{(1.04)}$&0.0 (1.59)\\
AND & 200 & 500 & $\bm{1.4}$ $\bm{(0.45)}$&0.0 (0.78)\\
\bottomrule
\end{tabular}%
\end{table}

\begin{table}[htbp]
\centering
\caption{Similar to Table~\protect\ref{sigma_90} but for the $95\%$ confidence intervals of $\sigma$.}
\label{sigma_95}
\begin{tabular}{l|ll|ll}
\toprule
function & $n$ & $p$& FART  &BART \\
\midrule
Cosine & 50 & 2 & $\bm{96.4}$ $\bm{(0.83)}$&15.5 (0.80)\\
Cosine & 200 & 2 & 99.1 (0.31)&$\bm{92.3}$ $\bm{(0.24)}$\\
XOR & 50 & 50 & $\bm{81.4}$ $\bm{(0.82)}$&9.6 (1.24)\\
XOR & 200 & 50 & $\bm{95.4}$ $\bm{(0.29)}$&96.1 (0.76)\\
AND & 50 & 500 & $\bm{26.1}$ $\bm{(1.27)}$&0.0 (1.90)\\
AND & 200 & 500 & $\bm{2.0}$ $\bm{(0.53)}$&0.0 (0.94)\\
\bottomrule
\end{tabular}%
\end{table}

\begin{table}[htbp]
\centering
\caption{Similar to Table~\protect\ref{sigma_90} but for the $99\%$ confidence intervals of $\sigma$.}
\label{sigma_99}
\begin{tabular}{l|ll|ll}
\toprule
function & $n$ & $p$& FART  &BART \\
\midrule
Cosine & 50 & 2 & $\bm{99.7}$ $\bm{(1.14)}$&36.3 (1.06)\\
Cosine & 200 & 2 & 99.9 (0.40)&$\bm{98.7}$ $\bm{(0.32)}$\\
XOR & 50 & 50 & $\bm{93.6}$ $\bm{(1.14)}$&24.0 (1.62)\\
XOR & 200 & 50 & 98.2 (0.38)&$\bm{98.9}$ $\bm{(1.07)}$\\
AND & 50 & 500 & $\bm{29.9}$ $\bm{(1.76)}$&0.1 (2.51)\\
AND & 200 & 500 & $\bm{3.5}$ $\bm{(0.70)}$&0.0 (1.25)\\
\bottomrule
\end{tabular}%
\end{table}

\begin{figure}[ht]
\begin{center}
\includegraphics[scale=0.3]{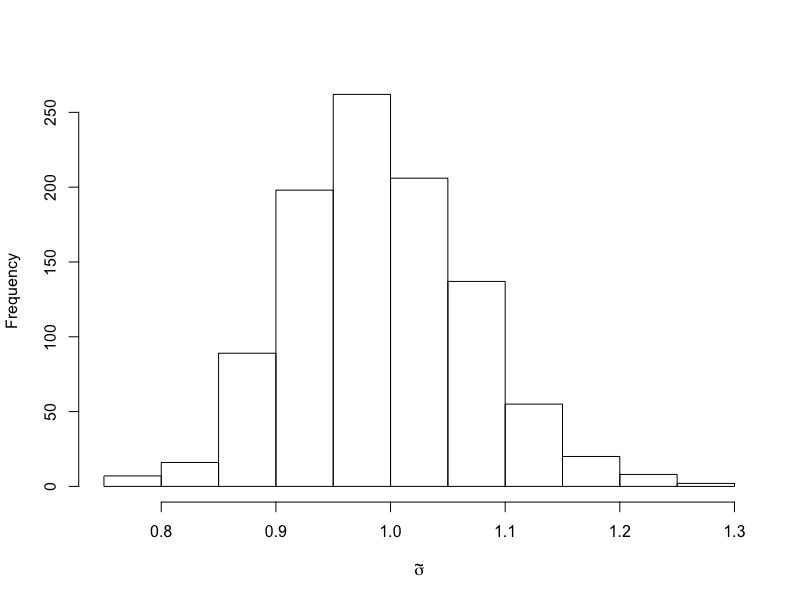}	
\caption{Histogram of the generalized fiducial samples $\tilde{\sigma}$ of $\sigma$.}
\label{sigma_hist}
\end{center}
\end{figure}

Lastly we provide the histogram of the generalized fiducial samples of $\sigma$, which can be seen as an approximation of the marginal generalized fiducial density of $\sigma$.  The histogram is displayed in Figure~\ref{sigma_hist}.  These samples were for the case when the test function is XOR with $n=200, p=50$.  One can see that the histogram is approximately bell-shaped and centered at the true value of $\sigma=1$.

\clearpage

\subsection{Real Data Examples}
\label{sec_real_data}
\begin{figure}
\begin{center}
\includegraphics[scale=0.3]{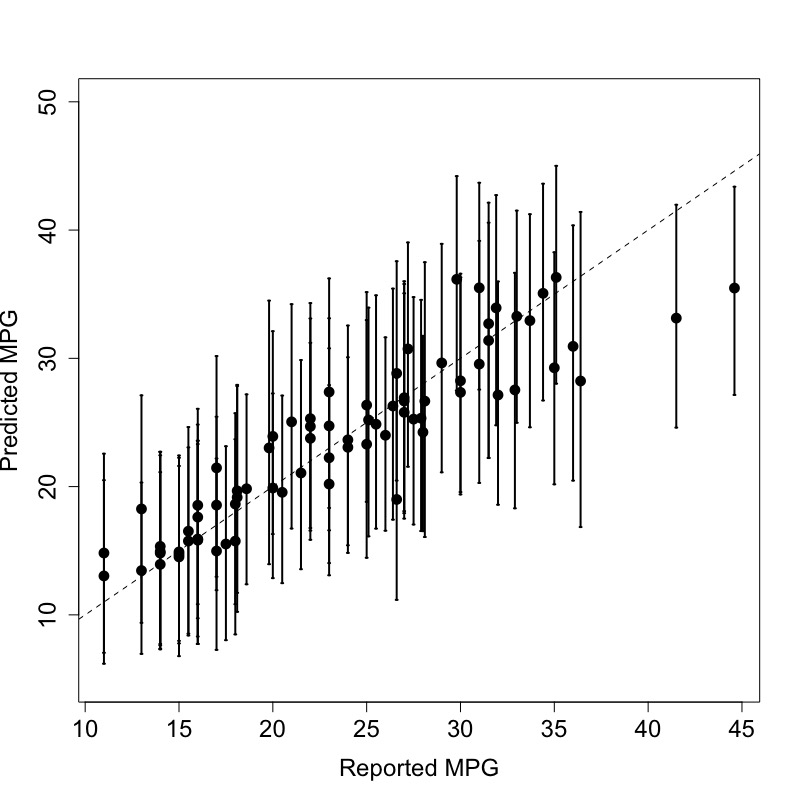}	
\caption{FART predictions and $95\%$ prediction intervals for the \emph{Auto MPG} data set.}
\label{auto_mpg2}
\end{center}
\end{figure}

This subsection reports the coverage rates the FART prediction intervals on five real data sets:
\begin{itemize}
\item \emph{Air Foil}: This is a NASA data set, obtained from a series of aerodynamic and acoustic tests of two and three-dimensional airfoil blade sections conducted in an anechoic wind tunnel \citep{Dua:2019}. Five features were selected to predict the aerofoil noise.  We used $1000$ observations as the training data set and $503$ observations as test data.
\item \emph{Auto Mpg}: This data set contains eight features to predict city-cycle fuel consumption in miles per gallon \citep{Asuncion+Newman:2007,Dua:2019}.  After discarded samples with missing entries, we split the rest of the observations into a training set of size $314$ and a test set of size $78$.
\item \emph{CCPP}: This data set contains $9568$ data points collected from a Combined Cycle Power Plant over six years (2006-2011), when the power plant was set to work with full load \citep{tufekci2014prediction,kaya2012local}.  There are four features aiming to predict the full load electrical power.  We split the data into a training set of size $8000$ and a test set of size $1568$.
\item \emph{Boston House}: Originally published by \citep{harrison1978hedonic}, a collection of $506$ observations associated with $14$ features from  U.S. Census Service are used to predict the median value of owner-occupied homes.  We split the data into a training set of size $400$ and a test set of size $106$.
\item \emph{CCS}: In civil engineering, concrete is the most important material \citep{yeh1998modeling}. This data set consists of eight features to predict the concrete compressive strength.  We split it into a training set of size $750$ and a test set of size $280$.	
\end{itemize}
For each of the above data sets, we applied FART to the training data set to construct 95\% prediction intervals for the observations in the test data set.  We repeated this procedure 100 times by randomly splitting the whole data set into a training data set and a test data set.  The empirical coverage rates of these prediction intercals are reported in Table~\ref{real_data}.  In addition, as a comparison to Figure~\ref{auto_mpg}, we plotted the coverage of the FART prediction intervals on the same \emph{Auto MPG} data in Figure~\ref{auto_mpg2}.  One can see that FART gave very good performances.  
\begin{table}[htbp]
\centering
\caption{Empirical coverage rates for the $95\%$ FART prediction intervals for various real data sets.  The numbers in parentheses are the averaged widths of the intervals.}
\label{real_data}
\begin{tabular}{l|ccccc}
\toprule
Data & \emph{Air Foil} & \emph{Auto Mpg} & \emph{CCPP}  & \emph{Boston House} & \emph{CCS}\\
\midrule
Coverage & 93.2\% (14.2) & 91.8\% (11.5) &95.1\% (11.5)& 87.9\% (12.4)& 92.8\% (30.4)\\
\bottomrule
\end{tabular}%
\end{table}

\section{Conclusion}
\label{sec:conclude}
In this paper, we applied generalized fiducial inference to ensembles of honest regression trees. In particular, we derived a fiducial probability for each honest tree in an honest random forests, which shows how likely the tree contains the true model. A practical procedure was developed to generate fiducial samples of the tree models, variance of errors and predictions.  These samples can further be used for point estimation, and constructing confidence intervals and prediction intervals. The proposed method was shown to enjoy desirable theoretical properties, and compares favorably with other state-of-the-art methods in simulation experiments and real data analysis.

\appendix
\section{Technical Details}
The appendix provides the proof for Theorem~\ref{t1}.

WLOG, assume $\sigma^2=1$ and fix $T\in\mathcal{T}_{l_0}$. We first prove that 
	\begin{equation*}
	\max_{T^{'}\notin\mathcal{T}_{l_0}, T^{'}\in \mathcal{T}}R(T^{'})/R(T) \rightarrow_p 0.
	\end{equation*}
	
	Rewrite
	\begin{equation*}
	R(T^{'})/R(T)=exp\{-D_1-D_2\},
	\end{equation*}
	where
	\begin{equation*}
	D_1=\frac{n-l(T^{'})-1}{2}\log\frac{\text{SSE}_{T^{'}}}{\text{SSE}_{T}},
	\end{equation*}
	\begin{equation*}
	D_2=\log\frac{\Gamma(\frac{n-l_0}{2})}{\Gamma(\frac{n-l(T^{'})}{2})}+\frac{l_0-l(T^{'})}{2}\log\pi+\frac{l_0-l(T^{'})}{2}\log \text{SSE}_{T}+\frac{l(T^{'})-l_0}{2}\log(n).
	\end{equation*}
	
{\bf Case 1:} $T^{'} \notin \mathcal{T}_0$.

Now calculate
\begin{equation}\label{eq2}
\begin{split}
\mbox{SSE}_{T^{'}}-\mbox{SSE}_{T}&=\Delta_{T^{'}}+2\bm{\mu}^{'}(\bm{I}-\bm{H}_{T^{'}})\bm{\varepsilon}-\bm{\varepsilon}^{'}(\bm{H}_{T^{'}}-\bm{H}_{T})\bm{\varepsilon}.
\end{split}
\end{equation}
Let $c_{l(T)}=l(T)\log\log n$, consider the second term in equation \eqref{eq2} and denote $Z_{T^{'}}=\bm{\mu}^{'}(\bm{I}-\bm{H}_{T^{'}})\bm{\varepsilon}/\sqrt{\Delta_{T^{'}}}$, then
\begin{equation*}
\mu^T(\bm{I}-\bm{H}_{T^{'}})\bm{\varepsilon}=\sqrt{\Delta_{T^{'}}}Z_{T^{'}}
\end{equation*}
and $Z_{T^{'}}\sim N(0,I_n)$ since $var(Z_{T^{'}})=1$.  Furthermore,
\begin{equation*}
\begin{split}
P(\mathop{\max}_{{T^{'}}\in \mathcal{T}}|Z_{T^{'}}/\sqrt{c_{l({T^{'}})}}|>1) 
& \leq |\mathcal{{T}}|\mathop{\max}_{{T^{'}}\in \mathcal{T}}P(Z_{T^{'}}^2>c_{l({T^{'}})})\\
&=|\mathcal{T}|\mathop{\max}_{{T^{'}}\in \mathcal{T}}P(\chi_1^2>c_{l({T^{'}})})\\
&\leq |\mathcal{T}|\mathop{\max}_{{T^{'}}\in \mathcal{T}}(c_{l({T^{'}})}e^{1-c_{l({T^{'}})}})^{1/2}\longrightarrow 0 \qquad as \quad n\longrightarrow \infty.
\end{split}
\end{equation*}
Therefore, $P(|\bm{\mu}^{'}(\bm{I}-\bm{H}_{T^{'}})\bm{\varepsilon}|>\sqrt{\Delta_{T^{'}}c_{l({T^{'}})}})\longrightarrow 0 \qquad as \quad n\longrightarrow \infty$.\\

Consider the third term in equation \eqref{eq2}:\\

Notice that $\bm{\varepsilon}^{'}\bm{H}_{T}\bm{\varepsilon}=\sum_{i=1}^{l(T)}n_i\bar{\epsilon}_i^2\sim\chi^2_{l(T)}$. Thus, \\
\begin{equation*}
\begin{split}
P(\mathop{\max}_{T\in\mathcal{T}}\bm{\varepsilon}^{'}\bm{H}_T\bm{\varepsilon}/c_{l(T)}> 1) & \leq |\mathcal{T}|\mathop{\max}_{T\in\mathcal{T}}P(\bm{\varepsilon}^T\bm{H}_{l(T)}\bm{\varepsilon} > c_{l(T)})\\
& =|\mathcal{T}|\mathop{\max}_{T\in\mathcal{T}}P(\chi_{l(T)}^2 > c_{l(T)})\\
& \leq |\mathcal{T}|\mathop{\max}_{T\in\mathcal{T}}(\frac{c_{l(T)}}{{l(T)}}e^{1-\frac{c_{l(T)}}{l(T)}})^{{l(T)}/2}\\
&=|\mathcal{T}|\mathop{\max}_{T\in\mathcal{T}}{(\frac{e\log\log n}{\log n})} ^{l(T)/2} \longrightarrow 0 \qquad as \quad n\longrightarrow \infty.
\end{split}
\end{equation*}
Therefore, $P(\bm{\varepsilon}^T\bm{H}_{T}\bm{\varepsilon}>c_{l(T)})\longrightarrow 0$, and
$P(\bm{\varepsilon}^T\bm{H}_{T^{'}}\bm{\varepsilon}>c_{l(T^{'})})\longrightarrow 0$ as $n\longrightarrow \infty$.
Thus, we have $P(\mbox{SSE}_{T^{'}}-\mbox{SSE}_{T}<0.5\Delta_{T^{'}})\longrightarrow 0$ as $n\longrightarrow \infty$.\\

In addition,
\begin{equation*}
	P(\chi_{n-L}^2<\frac{n}{4})\leq P(\chi_{n-L}^2<\frac{n-L}{2})\leq(\frac{\sqrt{e}}{2})^{\frac{n-L}{2}}\longrightarrow 0 \qquad as \quad n\longrightarrow \infty,
	\end{equation*}
	which means
	\begin{equation*}
	P(\min_{T^{'}\in \mathcal{T}}\chi_{n-l(T^{'})}^2<\frac{n}{4})\longrightarrow 0 \qquad as \quad n\longrightarrow \infty.
	\end{equation*}\\
Thus,
\begin{equation*}
\begin{split}
D_1&=\frac{n-l(T^{'})-1}{2}\log(\frac{\mbox{SSE}_{T^{'}}}{\mbox{SSE}_{T}})\\
&=-\frac{n-l(T^{'})-1}{2}\log(\frac{\mbox{SSE}_{T}}{\mbox{SSE}_{T^{'}}})\\
&=-\frac{n-l(T^{'})-1}{2}\log(1+\frac{\mbox{SSE}_{T}-\mbox{SSE}_{T^{'}}}{\mbox{SSE}_{T^{'}}})\\
&\geq \frac{n-l(T^{'})-1}{2}\frac{\mbox{SSE}_{T^{'}}-\mbox{SSE}_{T}}{\mbox{SSE}_{T^{'}}}\\
&=\Omega_p(\Delta_{T^{'}}).
\end{split}
\end{equation*}

Moreover, $D_2=\Omega_p(-l(T^{'})\log(n))$. Therefore, $D_1+D_2=\Omega_p(\log n)$.\\
	
	{\bf Case 2 :} $T^{'}\in \mathcal{T}_0$ and $l(T^{'})>l_0$.\\
	
	Recall $T\in\mathcal{T}_{l_0}$ is fixed. First notice that $\text{SSE}_{T}-\text{SSE}_{T'}$=$\chi^2_{l(T^{'})-l_0}(T^{'})$, where $\chi^2_{l(T^{'})-l_0}(T^{'})$ is a chi-square random variable depending on $T^{'}$ with degrees of freedom $l(T^{'})-l_0$.
	\begin{equation*}
	\begin{split}
	P(\max_{T^{'}\in \mathcal{T}_0,l(T^{'})>l_0}\frac{\chi^2_{l(T^{'})-l_0}(T^{'})}{(l(T^{'})-l_0)\log\log n}\geq 1)&\leq|\mathcal{T}| \max_{T^{'}\in \mathcal{T}_0,l(T^{'})>l_0}(\log\log ne^{1-\log\log n})^{\frac{l(T^{'})-l_0}{2}} \\
	&=|\mathcal{T}|(\frac{e\log\log n}{\log n})^{\frac{1}{2}}	\rightarrow 0.\\	
	\end{split}
	\end{equation*}
	It implies that 
	\begin{equation*}
	\chi^2_{l(T^{'})-l_0}= O_p(c_{l(T^{'})-l_0}).
	\end{equation*}
	
	Therefore,
	\begin{equation*}
	\begin{split}
	\frac{n-l(T^{'})-1}{2}\log\frac{\text{SSE}_{T^{'}}}{\text{SSE}_T}&=-\frac{n-l(T^{'})-1}{2}\log(1+\frac{\chi^2_{l(T^{'})-l_0}(T^{'})}{\chi^2_{n-l(T^{'})}})\\
	&\geq-\frac{n-l(T^{'})-1}{2}(\frac{\chi^2_{l(T^{'})-l_0}(T^{'})}{\chi^2_{n-l(T^{'})}})\\
	&=\Omega_p(-c_{l(T^{'})-l_0}),\\
	\end{split}
	\end{equation*}
	uniformly over $\{T^{'}:T^{'}\in \mathcal{T}_0$, $l(T^{'})>l_0\}$.
	Thus, we show that 
	\begin{equation*}
	D_1=\Omega_p(-\frac{l(T^{'})}{2}\log\log n).
	\end{equation*}
	
	Meanwhile, the calculation of $D_2$ is similar to Case 1, $D_2=\Omega_p((l(T^{'})-l_0)\log(n))$, so we have $D_1+D_2=\Omega_p(\log n)$.\\
	
	Combining Case 1 and Case 2, we have:
	\begin{equation*}
	\max_{T^{'}\notin\mathcal{T}_{l_0}, T^{'}\in \mathcal{T}}R(T^{'})/R(T)=O_p(1/n).
	\end{equation*}
	
	Furthermore, 
	\begin{equation*}
	\sum_{T^{'}\notin\mathcal{T}_{l_0}, T^{'}\in\mathcal{T}}R(T^{'})/R(T)\leq|\mathcal{T}| \max_{T^{'}\notin\mathcal{T}_{l_0}, T^{'}\in \mathcal{T}}R(T^{'})/R(T)\leq\frac{|\mathcal{T}|}{n} \rightarrow_p 0.
	\end{equation*}
	
	Equivalently, 
	\begin{equation*}
	\sum_{T\in \mathcal{T}_{l_0}}r(T)\rightarrow_p 1.
	\end{equation*}
\bibliographystyle{rss}
\bibliography{references}
\end{document}